\documentclass[twoside,11pt]{article}

\let\bibstyle\bibliographystyle 
\def\bibliographystyle#1{}      
\bibstyle{apalike-url}
\usepackage{obs_study_style}

\usepackage{graphicx} 
\usepackage{amsmath}
\usepackage{amssymb}
\usepackage{subcaption}
\usepackage{hyperref}
\usepackage{setspace}
\usepackage{enumerate}
\usepackage{xcolor}
\usepackage{dsfont}

\newcommand{\Soeren}{S\"oren }
\newcommand{\Kuenzel}{K\"unzel }

\newcommand{\ind}[1]{\mathds{1}(#1)}%

\heading{5}{2019}{21-35}{7/19}{7/19}{Carlos Carvalho, Avi Feller, Jared Murray, Spencer Woody, and David Yeager }

\ShortHeadings{Data Challenge: Assessing Treatment Effect Variation}{Carvalho, Feller, Murray, Woody, and Yeager} \firstpageno{1}

\begin{document}


\title{Assessing Treatment Effect Variation in Observational~Studies:
Results from a Data Challenge}

\author{\name Carlos Carvalho \email
  carlos.carvalho@mccombs.utexas.edu
  \addr 
  \\ Department of
  Information, Risk and Operations Management  \\
  The University of Texas at Austin\\
  Austin, TX 78712, USA
  \AND
  \name Avi Feller \email
  afeller@berkeley.edu \\
  \addr Goldman School of Public Policy  \\
  The University of California, Berkeley \\
  Berkeley, CA 94720, USA
  \AND
  \name Jared Murray \email
  jared.murray@mccombs.utexas.edu \\
  \addr Department of
  Information, Risk and Operations Management  \\
  The University of Texas at Austin\\
  Austin, TX 78712, USA
  \AND
  \name Spencer Woody  \email
  spencer.woody@utexas.edu \\
  \addr Department of Statistics and Data Science  \\
  The University of Texas at Austin\\
  Austin, TX 78712, USA
    \AND
    \name David Yeager \email
  dyeager@utexas.edu \\
  \addr Department of
  Psychology  \\
  The University of Texas at Austin\\
  Austin, TX 78712, USA
}

\maketitle


\begin{abstract}
  A growing number of methods aim to assess the challenging question of treatment effect variation in observational studies.  This special section of \emph{Observational Studies} reports the results of a workshop conducted at the 2018 Atlantic Causal Inference Conference designed to understand the similarities and differences across these methods.  We invited eight groups of researchers to analyze a synthetic observational data set that was generated using a recent large-scale randomized trial in education.  Overall, participants employed a diverse set of methods, ranging from matching and flexible outcome modeling to semiparametric estimation and ensemble approaches.  While there was broad consensus on the topline estimate, there were also large differences in estimated treatment effect moderation.  This highlights the fact that estimating varying treatment effects in observational studies is often more challenging than estimating the average treatment effect alone.  We suggest several directions for future work arising from this workshop.
\end{abstract}



\section{Introduction}
\label{sec:introduction}

Spurred by recent statistical advances and new sources of data, a growing number of methods aim to assess treatment effect variation in observational studies.
This is an inherently challenging problem.  Even estimating a single overall effect in non-randomized settings is difficult, let alone estimating how effects vary across units.
As applied researchers begin to use these tools, it is therefore important to understand both how well these approaches work in practice and how they relate to each other.

This special section of \emph{Observational Studies} reports the results of a workshop conducted at the 2018 Atlantic Causal Inference Conference designed to address these questions.
Specifically, we invited researchers to analyze a common data set using their preferred approach for estimating varying treatment effects in observational settings.
The synthetic data set, which we describe in detail in Section \ref{sec:datagen}, was based on  data from the National Study of Learning Mindsets, a large-scale randomized trial of a behavioral intervention \citep{nslm}. Unlike the original study, the simulated dataset was constructed to include meaningful \emph{measured} confounding, though the assumption of no unmeasured confounding still holds.
We then asked participants to answer specific questions related to treatment effect variation in this simulated data set and to present these results at the ACIC workshop.

Workshop participants submitted a total of eight separate analyses.
At a high level, all contributed analyses followed similar two-step procedures: (1) use a flexible approach to unit-level treatment effects; (2) find low-dimensional summaries of the estimates from the first step to answer the substantive questions of interest.
The analyses, however, differed widely in their choice of flexible modeling, including matching, machine learning models, semi-parametric methods, and ensemble approaches.
In the end, there was broad consensus on the topline estimate, but large differences in estimated treatment effect moderation.
This underscores that estimating varying treatment effects in observational studies is more challenging than estimating the ATE alone.
Section~\ref{sec:overview} gives an overview of the data challenge and the proposed methods. Section~\ref{sec:summary-findings} discusses the contributed analyses. Section~\ref{sec:common-themes} addresses common themes and highlights some directions for future research. Participants' analyses appear subsequently in this volume.
\setcounter{page}{22}

\section{Overview of the data challenge}\label{sec:overview}

\subsection{Background and problem setup}
The basis for the data challenge is the National Study of Learning
Mindsets \citep{nslm, Yeager2019}, which several workshop organizers
helped to design and analyze.  NSLM is a large-scale randomized
evaluation of a low-cost ``nudge-like'' intervention designed to
instill students with a \emph{growth mindset}.  At a high level, a
growth mindset is the belief that people can develop intelligence, as
opposed to a \emph{fixed mindset}, which views intelligence as an
innate trait that is fixed from birth.  NSLM assessed this
intervention by randomizing students separately within 76 schools
drawn from a national probability sample of U.S. public high schools.
In addition to assessing the overall impact, the study was designed to
measure impact variation, both across students and across schools. See
\citet{nslm} for additional discussion.

The goal in generating the synthetic data was to create an observational study that emulated NSLM in key ways, including covariate distributions, data structures, and effect sizes, but that also introduced additional confounding not present in the original randomized trial. The final dataset included 10,000 students across 76 schools, with four student-level covariates and six school-level covariates shown in Table~\ref{tab:data-description}. These covariates were drawn from the original National Study but were slightly perturbed to ensure privacy and other data restrictions (see Appendix~\ref{sec:dataapp} for details). For each student, we then generated a simulated outcome $Y$, representing a continuous measure of achievement, and a simulated binary treatment variable $Z$. We describe the data generating process in Section~\ref{sec:datagen}, with details in Appendix~\ref{sec:dataapp}.

\begin{singlespacing}
    \begin{table}[tb]
      \centering
      \begin{tabular}[ht]{ | l | p{0.5\textwidth} | } \hline Covariate
           & Description                                                                                                                        \\ \hline
        S3 & Student's self-reported expectations for success in the future, a proxy for prior achievement, measured prior to random assignment \\ \hline
        C1 & Categorical variable for student race/ethnicity                                                                                    \\ \hline
        C2 & Categorical variable for student identified gender                                                                                 \\ \hline
        C3 & Categorical variable for student first-generation status (i.e., first in family to go to college)                                  \\ \hline \hline
        XC & School-level categorical variable for urbanicity of the
                      school (i.e., rural, suburban, etc.)                                                                                      \\ \hline
        X1 & School-level mean of students' fixed mindsets, reported
                         prior to random assignment                                                                                             \\ \hline
        X2 & School achievement level, as measured by test scores and
                         college preparation for the previous four cohorts of students                                                          \\ \hline
        X3 & School racial/ethnic minority composition -- i.e.,
                         percentage black, Latino, or Native American                                                                           \\ \hline
        X4 & School poverty concentration -- i.e., percentage of
                         students who are from families whose incomes fall below the federal
                         poverty line                                                                                                           \\ \hline
        X5 & School size -- Total number of students in all four grade
                         levels in the school                                                                                                   \\ \hline
      \end{tabular}
      \caption{Descriptions of available covariates in the ACIC workshop synthetic dataset}
      \label{tab:data-description}
    \end{table}

\end{singlespacing}

Participants were presented with several objectives for their analysis:

\begin{enumerate}
\item To assess whether the mindset intervention is effective in improving student achievement
\item To assess two potential effect moderators of primary scientific interest: Pre-existing mindset norms (X1) and school level achievement (X2). In particular, participants were asked to evaluate two competing hypotheses about how X2 moderates the effect of the intervention: if it is an effect modifier, researchers hypothesize that either it is largest in middle-achieving schools (a ``Goldilocks effect'') or is decreasing in school-level achievement.
\item To assess whether there are any other effect modifiers among the recorded variables.
\end{enumerate}

We chose these objectives in part because they arose in the design and analysis of the original National Study. We intentionally kept the wording statistically vague, and did not map the objectives onto specific estimands in order to emulate part of our experience working with collaborators.

\subsection{Overview of data generation}\label{sec:datagen}

According to the model generating synthetic data:
\begin{enumerate}
	\item The mindset intervention has a relatively large, positive effect on average.
	
	\item The impact varies across both pre-existing mindset norms (X1) and school-level achievement (X2). However, there is no ``Goldilocks'' effect present, at least when controlling for other variables (C1 and X1).
	
	\item The impact also varies across race/ethnicity (C1).
\end{enumerate}
Where possible we anchored the synthetic data generating process to the original data, borrowing the original covariate distribution (with slight perturbations) and using semiparametric models fit to an immediate post-treatment outcome from the original study.

Specifically, let $w_{ij}$ denote the vector of the variables in Table~\ref{tab:data-description} for student $i$ in school $j$. We generated the data from the following model:
\begin{equation}\label{eq:sim_dgp}
y_{ij}^{\text{obs}} = \alpha_j + \mu(w_{ij}) + [\tau(x_{j1}, x_{j2}, c_{ij1}) + \gamma_j]z_{ij} + \epsilon_{ij},
\end{equation}
where $\mu$ is an additive function obtained approximately by fitting a generalized additive model to the control arm of the original data; see Appendix~\ref{sec:dataapp}. We simulated $\alpha_j\sim N(0, 0.15^2)$ and $\gamma_j\sim N(0, 0.105^2)$ independently. We drew iid samples of $\epsilon_{ij}$  by jittering and resampling the residuals from a model fit to the original data, scaling them to have standard deviation 0.5; the distribution of the error terms is displayed in Figure~\ref{fig:partial}. Finally, we generated treatment effects from the following model:
\begin{equation}\label{eq:tau_dgp}
\tau(x_1, x_2, c_1) = 0.228 + 0.05\cdot\ind{x_1<0.07} - 0.05\cdot \ind{x_2<-0.69} - 0.08\cdot \ind{c_1\in \{1, 13, 14\}},
\end{equation}
where $x_1$ is a measure of pre-existing mindset norms, $x_2$ is school-level achievement, and $c_1$ is a categorical race/ethnicity variable. All three appeared to be associated with treatment effect variation in preliminary data from the original National Study, although we adjusted the pattern for the synthetic dataset.
Note that there is no additional idiosyncratic treatment effect variation and the control and treatment potential outcomes have the same (conditional) variance.

We generated confounding via a two-step process. First, we dropped observations from the treatment arm with probability ${1-\Phi\left[-0.5+1.5\mu(w_{ij})\right]}$, where $\Phi$ is the standard normal CDF. This simulates a scenario where students with high expected outcomes under control were more likely to receive the treatment, yielding naive treatment effect estimates that are too high.
Second, similar to the design in \citet{Hill2011}, we dropped select units from the treatment arm to induce a more complicated functional form for the confounding structure.  Specifically, we dropped students from the treatment arm if they were above the 80\textsuperscript{th} percentile on an additional covariate from the National Study that was not included in the synthetic dataset (and not used in generating the synthetic outcomes).  In principle this has the potential to induce violations of overlap in the presence of very strong dependence, but overlap in the final dataset was quite good (see e.g. Carnegie et al and Johannsson for overlap checks).

As with any synthetic data simulation, the process involved an extensive number of modeling choices. Appendix~\ref{sec:dataapp} contains more details about the data generating process and discusses the principles we formulated prior to generating the data, including decisions about the relative magnitude of the average treatment effect, treatment effect heterogeneity, and residual variance.

\section{Overview of contributed analyses}\label{sec:summaryfindings}
\label{sec:summary-findings}

\subsection{Summary of contributed methods}

Participants submitted eight separate analyses using a wide range of methods.\footnote{Alejandro Schuler participated in the ACIC workshop but did not submit a written analysis.}
At a high level, all of the contributed analyses followed similar two-step procedures: (1) use a flexible approach to impute student-level (or school-level) treatment effects; (2) find low-dimensional summaries of the flexible model to answer the substantive questions. Specifically, the proposed first-stage methods fall into three broad categories:
\begin{itemize}

	\item \emph{Matching.} Keller et al; Keele and Pimentel; and Parikh et al;
		
	\item \emph{Outcome modeling.} Carnegie et al.; Johannsson;
	
	\item \emph{Machine learning and semi-parametric estimation.} Zhao \& Panigrahi; Athey \& Wager.
\end{itemize}

 \noindent \Kuenzel et al. proposed an ensemble approach that could incorporate any number of candidate estimators.
Approaches for the second step range from graphical summaries and simple aggregation (e.g., averaging student-level impact estimates by school) to fitting regression or tree models to the estimated student-level impacts.

\subsection{Summary of findings}
At a high level, most analyses resulted in remarkably similar estimates for the average treatment effect.  Conversely, the proposed methods generally do not agree on treatment effect heterogeneity, though
the presented results rarely contradict each other: While some methods reported discovered effect modifiers that others did not detect, no two methods, for example, found effects varying in opposite directions.

\paragraph{Objective 1: average effect.}
Table~\ref{tab:ATE-summary} shows the reported estimates of the Average Treatment Effect and corresponding 95\% uncertainty intervals.\footnote{The original prompt was intentionally vague on the target estimand. Most respondents reported estimates for an Average Treatment Effect. Keele \& Pimentel specifically reported estimates for an Average Treatment Effect on the Treated. Athey \& Wager reported an Average Treatment Effect for a population of sites. In the end, the true values are quite similar across these estimands in the synthetic data set. Finally, all uncertainty intervals are confidence intervals, except for Carnegie et al., who report credible intervals.}
Overall the point estimates are quite similar, ranging from 0.25 to 0.27, compared to a true value of 0.24.
The interval widths varied widely, however, from 0.01 for Parikh et. al. to 0.09 for Johannsson.
And while most intervals bracketed the true value, with the exceptions of Parikh et. al. and Keele and Pimentel.\footnote{The true Sample Average Treatment Effect on the Treated was also about 0.24.}
we cannot assess the quality of these intervals based on a single realized synthetic data set.

These results are consistent with our own experiences and anecdotal evidence suggesting that --- in settings where the identifying assumptions hold, sample sizes are large, and the signal is reasonably strong --- estimates of the overall effect are fairly robust to modeling or analytic choices.
Inference may be another story, however. While all eight analyses point to large, positive effects overall, the variation in interval widths suggests differences in efficiency and/or frequentist validity of uncertainty intervals, or possibly differences in the target estimand; for example, Athey and Wager target the population {\em school} average treatment effect, while Keele and Pimentel estimate the sample ATT. 

\begin{table}[tb]
  \centering
  \begin{tabular}[ht]{|l|r|}
    \hline
    Author            & ATE estimate (95 \% C.I.) \\ \hline
    Athey \& Wager    & 0.25 (0.21, 0.29)                     \\
    Carnegie et al.   & 0.25 (0.23, 0.27)                     \\
    Johannsson        & 0.27 (0.22, 0.31)                     \\
    Keele \& Pimentel & 0.27 (0.25, 0.30)                     \\
    Keller et al.     & 0.26 (0.22, 0.30)                     \\
    \Kuenzel et al.   & 0.25 (0.22, 0.27)                     \\
    Parikh et al.     & 0.26 (0.25, 0.26)                     \\
    Zhao \& Panigrahi & 0.26 (0.24, 0.28)                     \\ \hline
  \end{tabular}
  \caption[]{
  Submitted estimates for the average treatment effect and corresponding 95\% uncertainty intervals.
  }
  \label{tab:ATE-summary}
\end{table}

\begin{table}[tb]
  \centering
  \begin{tabular}{|l|p{5cm}|p{5cm}|}
    \hline
    Author            & School mindset norms (X1)                 & School achievement level (X2)                    \\ \hline
    Athey \& Wager    & \textbf{Yes}, though negative effect modification is insignificant after multiplicity adjustment.    & \textbf{No.}  \\ \hline
    Carnegie et al.   & \textbf{Yes}, decreasing trend in treatment effect. & \textbf{Yes}, an increasing trend in treatment effect, though without evidence of a Goldilocks effect \\ \hline
    Johannsson        & \textbf{Yes}, decreasing trend in treatment effect. &     \textbf{No.}  \\ \hline
    Keele \& Pimentel & \textbf{Yes}, decreasing trend in treatment effect, though confidence intervals for quintiles overlap.  & \textbf{Moderate support}; lowest quintile has lower treatment effect, though there is little separation among other quintiles.  \\ \hline
    Keller et al.     & \textbf{Moderate support} from exploratory analysis.    &  \textbf{Moderate support} for presence Goldilocks effect. \\ \hline
    \Kuenzel et al.   & \textbf{Yes}. increasing trend in treatment effect.   & \textbf{No}, though there is cursory evidence in CATE plots. \\ \hline
    Parikh et al.     & \textbf{Yes}, increasing then decreasing trend in treatment effect.   &  \textbf{Yes}, exploratory support for Goldilocks effect.   \\ \hline
    Zhao \& Panigrahi & \textbf{Yes}, though negative effect modification is insignificant after selection adjustment.   & \textbf{No.} \\
    \hline
  \end{tabular}
  \caption{Conclusions about Objective 2 concerning effect modification by X1
    and X2}
  \label{tab:question-2}
\end{table}

\paragraph{Objective 2: variation across pre-specified moderators.} In contrast to responses concerning the overall effect, there is considerable disagreement about treatment effect variation across the two pre-specified moderators; see Table~\ref{tab:question-2}.
All participants found at least some support for effect modification across baseline levels of mindset beliefs (X1) and generally agreed on the direction: higher baseline mindset beliefs associated with lower treatment effects on average.  However, there was some disagreement on whether this trend should be considered ``statistically significant.'' As we discuss in Section~\ref{sec:common-themes}, this is due in part to different standards for evidence as well as challenges in quantifying uncertainty for some approaches. 

Results were more mixed for variation by baseline student achievement (X2). Half of the analyses found essentially no support for variation across X2. The remaining analyses were split as to the magnitude and strength of evidence for variation, though all were suggestive of an increasing trend.

\paragraph{Objective 3: exploratory treatment effect variation.} The contributed analyses also differed in their conclusions about additional effect modifiers. (Recall from Eq~\eqref{eq:sim_dgp} that in the data generating process, X1, X2, and C1 were all ``true'' treatment effect modifiers, with treatment effects increasing in X1, decreasing in X2, and smaller for C1=1,13, or 14) .

With the exception of Athey \& Wager, who found no additional treatment effect variation, the remaining seven analyses all identified urbanicity (XC) as an effect modifier, although they varied somewhat in their assessment of the strength of evidence. Some authors also identified student self-reported expectations for future success (S3) as a possible effect modifier, generally with an increasing trend.  These findings were interesting in part because neither XC nor S3 are ``true'' effect modifiers in the sense of appearing in the data generating process for $\tau$ in Equation~\eqref{eq:tau_dgp}, although XC is associated with $\tau$ in the population.

\begin{figure}
 \centering
 \includegraphics[width=.97\textwidth]{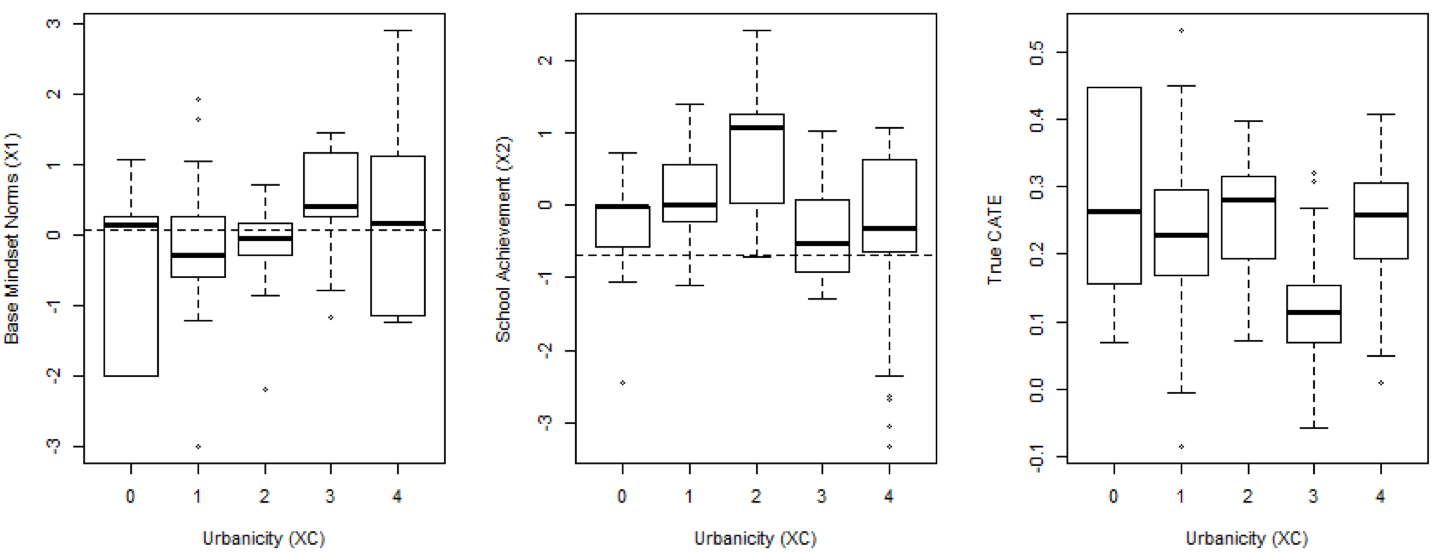}
\caption{Relationships between the Urbanicity variable (XC) and other quantities, from left to right: Base mindset norms (X1), school achievement (X2), and the true CATE. In the left two panels we see how XC correlates with X1 and X2.  The dotted lines are the locations of the step for the CATE function $\tau$ in Equation~\eqref{eq:tau_dgp} for each variable (X1 and X2). Note that, in addition to the sample correlation between X1/X2 and XC, observations with $\text{XC}=3$ tended to have norms (X1) above the X1 cutpoint in $\tau$ and achievement (X2) below the X2 cutpoint in $\tau$. In the last panel we see that {\em marginally} the true CATEs for observations with $\text{XC}=3$ are notably lower than the other levels.}
 \label{fig:XC}
\end{figure}

This highlights two important issues in assessing treatment effect variation in non-randomized studies.
First, even in a randomized trial, estimating varying treatment effects is, in some sense, an inherently observational problem and is generally susceptible to Simpson's paradox 
\citep{vanderweele2011interpretation}.
For example, urbanicity is strongly related to the true effect modifiers X1 and X2: if the analysis does not condition on X1 and X2, the estimated treatment effects will vary across levels of XC; see Figure \ref{fig:XC}.\footnote{The observed relationship between urbanicity and treatment is further amplified by the step function specification in Equation~\eqref{eq:tau_dgp} as well as the particular draw of random slopes, $\gamma_j$, in the realized synthetic data set.}
Therefore, whether urbanicity should be considered a ``true'' effect modifier depends on whether the analysis conditions on X1, X2, and C1, and on whether we are estimating heterogeneity across sampled schools or in the population. The research question given to workshop participants was intentionally vague on this point, and different authors interpreted the question in different ways.

\begin{figure}
 \centering
 \includegraphics[width=.97\textwidth]{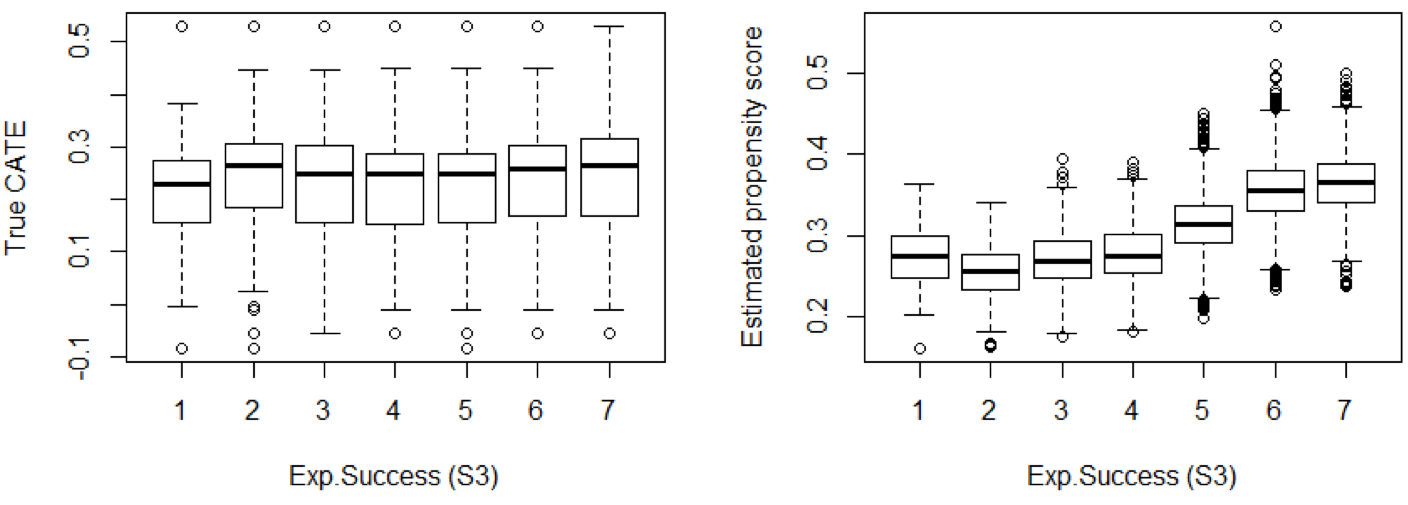}
\caption{True CATE and estimated propensity score by level of expected success (S3). Unlike XC, we see relatively little variability in CATE by S3 (left panel). However, S3 was one of the most important confounders (the right panel above shows how the propensity score varies with S3; Figure~\ref{fig:gam} shows that S3 is also very predictive of the outcome).}
 \label{fig:S3}
\end{figure}

Second, disentangling differential confounding from treatment effect variation is inherently challenging in observational settings. Unlike XC, S3 has essentially no relationship with the true individual-level treatment effects {\em unconditionally}, but is a very important predictor of selection into treatment.  In the analyses that found evidence for treatment effect heterogeneity in S3, the estimated pattern of heterogeneity lined up well with the patterns of confounding -- higher estimated treatment effects in the upper two levels of S3 -- suggesting that the estimated treatment effect variation may be due to residual confounding; see Figure~\ref{fig:S3}. These results are noteworthy in part because the submitted analyses were able to more or less correctly estimate the \emph{overall} treatment effect despite confounding, which underscores that estimating varying treatment effects in observational studies is more challenging than estimating the ATE.


\section{Common themes and open questions}\label{sec:common-themes}

\subsection{Making substantive questions statistically precise}
As with all real-world analyses, there are important open-ended
questions in terms of translating the substantive goals of the
research study into corresponding statistical methods.


One important challenge is weighing evidence for pre-defined subgroups versus exploratory treatment effect variation.
The proposed methods differ widely.
At one extreme, Zhao and Panigrahi have different inferential procedures for
pre-determined candidate subgroup effects (school-level fixed mindset
X1 and achievement level X2, as specified in Question 2) versus
``discovered'' subgroup effects, as mentioned in Question 3.
They argue that using the same data for both discovery and estimation of effect modification will generally lead to bias \citep{Fithian}, which motivates their post-selection inference procedure \citep{lee2016}.  
%
At the other extreme, Carnegie et al. treat all variation the same,
and conduct their investigation into treatment effect heterogeneity by
leveraging posterior uncertainty from BART. Most approaches lie
between these two extremes, such as using different sample splits to define subgroups versus estimating their impacts. (\Kuenzel et al.).

Another difference across methods is how submitted analyses addressed the multilevel structure of the data. Approaches included using ``cluster-robust'' random forests
(Wager \& Athey), bootstrapping and sample splitting at the school
level (Johannson; \Kuenzel et al.), including fixed or random intercepts
at the school level (Carnegie et al.), and considering matching both
between and within schools (Keele \& Pimentel).
These  approaches make different assumptions about patterns of treatment effect moderation as well as the estimand of interest. For example, Carnegie et. al.'s choice of random intercepts per school relaxes the assumption of iid errors at the student level but does not allow the effect of treatment effect moderators to vary across schools.
Alternatively, Athey \& Wager use a cluster-robust approach and an
estimator that targets a population of schools, rather than (as was
often implicitly targeted by the other analyses) a population of
students drawn from the 76 observed schools.

Third, as we mention in Section~\ref{sec:summaryfindings}, the contributed analyses all use a two-step approach for assessing treatment effect variation, first fitting a flexible model for impacts and then finding a low-dimensional summary of that model.
While a promising framework, the statistical properties of such procedures are not well understood.
First, as \citet{BCF} argue, this is generally valid from a Bayesian perspective, since it simply entails summarizing the posterior distribution after conditioning on the data only once.
The story, however, is more complicated from a frequentist perspective. \citet{chernozhukov2018generic} offer some promising directions in the randomized trial setting; see also the discussion in Athey \& Wager and Zhao \& Panigrahi. These ideas merit further exploration.


\subsection{Tailoring methods for observational studies}
The (simulated) data for this data challenge come from an
observational study rather than a randomized trial. Therefore,
appropriately accounting for confounding is a first-order
concern.  While all the analyses adjusted for confounding, the contributions varied in the level of emphasis put on the observational aspect of the data exercise, which creates additional challenges for estimating treatment effect variation.
%
In particular, when a method only partially
accounts for confounding, it is possible to confuse differential
confounding for varying treatment effects.  This appears to be driving some of the findings of S3 as an effect modifier, as we discuss in Section~\ref{sec:summaryfindings}.
 Understanding this dimension will be critical for broader adoption of these methods.

First, it is common to check for global covariate balance when estimating
overall impact in an observational study. A natural extension to
heterogeneous treatment effect estimation would be to also check for
\emph{local} covariate balance, such as separately by candidate
subgroups.  Even if we attain global covariate balance to a reasonable tolerance, imbalance within subgroups should give us pause. Such analyses were largely absent from the contributed
papers, suggesting that the field should develop standards in this
area.

Another element that is critical for observational studies is
assessing the consequences when the unconfoundedness assumption fails, either globally or locally.  Keele \&
Pimentel assess sensitivity within the matching framework; Carnegie et
al. instead use a model-based approach, and \Kuenzel et al. use a
permutation approach.
Sensitivity analysis for treatment effect variation remains an active research area, with some recent proposals within the minimax framework \citep{kallus2018confounding, yadlowsky2018bounds}.

\section{Conclusion}

Our goal for this workshop was to understand how different researchers would approach the kinds of questions we routinely face in our own applied work, as a complement to existing data analysis competitions like the ACIC data challenge \citep{dorie} where methods are formally evaluated based on their operating characteristics. We were fortunate to bring together a ``methodologically diverse'' panel and set them to analyze a single dataset, giving us a unique opportunity to compare perspectives. 
Our hope was that we would learn more about new methods while finding areas of overlap and points of divergence that suggest new lines of research.  We suggest a few directions for future work above, but fully expect this synthesis is the first word and not the last.  We sincerely thank the workshop participants and the other contributors to this volume for making the workshop such a success.



\acks{We gratefully acknowledge support from the Center for Enterprise
  and Policy Analytics at the McCombs School of Business at the
  University of Texas at Austin.  Preparation of this manuscript was
  supported by the William T. Grant Foundation, the National Science
  Foundation under grant number HRD 1761179, and by National Institute
  of Child Health and Human Development (Grant No. 10.13039/100000071
  R01HD084772-01, and Grant No. P2C-HD042849, to the Population
  Research Center [PRC] at The University of Texas at Austin). The
  content is solely the responsibility of the authors and does not
  necessarily represent the official views of the National Institutes
  of Health and the National Science Foundation.

}




\appendix

\section{Additional details for generating synthetic data}\label{sec:dataapp}
\subsection{Principles for data generation}

We had several goals for our data generating process.
Since we had a single dataset to generate based on a scientific problem with which we were quite familiar, we had the luxury of thinking carefully about plausible structures for treatment effect heterogeneity and confounding.

Our model for generating treatment effect heterogeneity was based on the following principles:

\begin{enumerate}
\item The average treatment effect should be relatively well-estimated by any reasonable procedure for confounding adjustment.
\item The variability in conditional average treatment effects (CATEs) should be relatively modest; here the covariate-dependent component of the CATEs ranged from 0.10 to 0.28, while the total CATEs (including school-level heterogeneity unexplained by the covariates) ranged from -0.08 to 0.53 and have a sample average of 0.24 and standard deviation 0.10. ``Relative'' here should be understood in terms of the marginal standard deviation of $Y$ (about 0.6) and the standard deviation of the error term in the model we used to generate the data (0.5).
\item It should be possible to approximately recover the treatment effect heterogeneity at conventional levels of statistical significance {\em given complete knowledge of the correct functional form}. This is an obvious baseline. We also tried to ensure that it was possible to recover at least some aspects of treatment effect moderation using plausible methods in a fully exploratory fashion, or when the true set of treatment effect moderators was known.
\item There should be no additional unmeasured treatment effect moderation at the individual level, since this is inestimable. This is wholly unrealistic and was primarily a variance-reduction decision; to the extent that other simulation exercises are explicit about this point, they tend to simulate unmeasured treatment effect moderation as due to independent unmeasured variables (e.g. \citet{dorie}). This is closer to reality of course, but in our context with a single dataset to be analyzed we saw little to be gained by injecting additional variability into the problem.
\item Unexplained treatment effect heterogeneity at the group level should be present at reasonable levels. We achieved this by simulating a random slope per school from a normal distribution with mean zero and standard deviation slightly larger than what we observed in models fit to the original data.
\end{enumerate}

Our confounding structure was based on the following principles:
\begin{enumerate}
\item The confounding should be strong enough to matter, but not so strong as to be unrealistic and/or induce practical violations of overlap. In our context this meant a naive, unadjusted estimate average treatment effect (ATE) estimate was about 25\% higher than the estimate using a correctly specified model (roughly, 0.30 unadjusted versus 0.24 adjusted, a difference of multiple standard errors).
\item The confounding should be explicitly modeled, at least in part, rather than induced solely by randomly sampling coefficients in outcome and selection models. In particular, we assumed that selection into treatment was partially based on expected outcomes. Here students with higher expected outcomes under control were more likely to be treated. Rather than violate conditional ignorability/unconfoundedness via a (perhaps more plausible) latent variable model, we made the selection model depend directly on $\mu_0(w) := \text{E}(Y(0)\mid W=w)$, where we use $W$ generically to denote the collection of potential effect moderators and confounders. \cite{BCF} refer to this confounding structure as ``targeted selection.''
\item We entertained inducing confounding at the group level that was unexplained by group-level covariates, but decided against it. This was largely a pragmatic decision, as it was already surprisingly difficult to design a data generating process that met our other desiderata.
\end{enumerate}

We wanted the covariate distribution to be realistic. We were interested in the practical effects of dependence in the covariates on estimating effect variation, and in how participants would interpret the research questions in light of this dependence.  We took the covariates almost directly from the early Mindset study data. To satisfy privacy constraints and avoid premature disclosure of variables constructed by the Mindset team, the original covariates were slightly perturbed. We added noise to continuous variables, where the disturbances were sampled from multivariate normal distributions that preserved the covariance structure (marginally over categorical variables). Categorical variables were subject to low levels of data swapping. In both cases we preserved the original multilevel data structure.

\subsection{Summary of original GAM estimate}

\begin{figure}
 \centering
 \includegraphics[width=.7\textwidth]{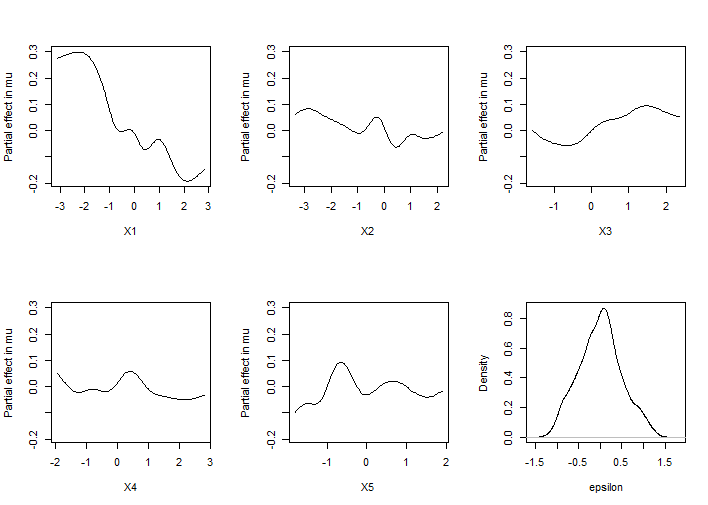}
\caption{Nonlinear terms in $\mu$ and error distribution (bottom right panel) for the data generating model.}
 \label{fig:partial}
\end{figure}

See Figure~\ref{fig:partial}.

\begin{figure}
 \centering
\begin{verbatim}
Parametric coefficients:
             Estimate Std. Error t value Pr(>|t|)
(Intercept) -0.279038   0.014847 -18.794  < 2e-16 ***
S3 == 2     -0.009130   0.014948  -0.611 0.541351
S3 == 3     -0.017494   0.013360  -1.309 0.190402
S3 == 4      0.046260   0.012716   3.638 0.000276 ***
S3 == 5      0.252158   0.012290  20.518  < 2e-16 ***
S3 == 6      0.614424   0.012315  49.893  < 2e-16 ***
S3 == 7      0.964031   0.012872  74.891  < 2e-16 ***
C1 == 2      0.011635   0.005728   2.031 0.042255 *
C1 == 3     -0.035327   0.012573  -2.810 0.004966 **
C1 == 4     -0.107783   0.004981 -21.638  < 2e-16 ***
C1 == 5      0.215135   0.008003  26.882  < 2e-16 ***
C1 == 6      0.001796   0.021055   0.085 0.932026
C1 == 7      0.034712   0.021602   1.607 0.108097
C1 == 8     -0.073001   0.010092  -7.233 4.96e-13 ***
C1 == 9     -0.138227   0.011785 -11.729  < 2e-16 ***
C1 == 10    -0.099958   0.010868  -9.197  < 2e-16 ***
C1 == 11    -0.084852   0.010809  -7.850 4.45e-15 ***
C1 == 12     0.049686   0.008453   5.878 4.25e-09 ***
C1 == 13    -0.100919   0.010573  -9.545  < 2e-16 ***
C1 == 14     0.064214   0.006531   9.832  < 2e-16 ***
C1 == 15    -0.019233   0.008204  -2.344 0.019078 *
C2          -0.155658   0.002546 -61.127  < 2e-16 ***
C3          -0.093056   0.002850 -32.655  < 2e-16 ***
XC           0.019555   0.002527   7.740 1.07e-14 ***
---

        edf Ref.df      F p-value
s(X1) 8.718  8.950 128.08  <2e-16 ***
s(X2) 8.920  8.987  25.08  <2e-16 ***
s(X3) 6.726  7.661  62.31  <2e-16 ***
s(X4) 7.827  8.554  19.73  <2e-16 ***
s(X5) 8.818  8.974  58.23  <2e-16 ***
---

\end{verbatim}
 \caption{GAM summary table of $\mu$ for the true expected outcomes
   under control, see Figure~\ref{fig:partial} for the forms of
   nonlinear terms in X1 through X5.}
 \label{fig:gam}
\end{figure}

\section{List of participants}
\label{sec:participants}

In total there were nine participants in the workshop, all of whom gave
presentations of their findings.  Here we list the presenters by order
of appearance, along with a brief description of their methodology:

\begin{singlespacing}
  \begin{itemize}
  \item Stefan Wager (Stanford): Causal random forests and the R-Learner
    transformed outcome
  \item Alejandro Schuler (Stanford): Shallow CART fit to transformed
    outcome (R-Learner).
  \item Luke Keele (Penn) and Sam Pimentel (UC-Berkeley): Various matching-based approaches.
  \item Qingyuan Zhao (Penn): Linear model on a transformed outcome
    and treatment with lasso regularization, with selection adjustment
    for inference about nonzero coefficients.
  \item \Soeren \Kuenzel (UC-Berkeley): Ensemble method to detect
    candidate subgroups, using standard ATE estimators within these
    subgroups.
  \item Nicole Carnegie (Montana State) and Jennifer Hill (New York
    University): BART with and without random intercepts.
  \item Alexander Volfovsky (Duke): Matching after learning to stretch
    (MALTS), a matching method that infers a distance metric.
  \item Frederik Johannsson (MIT): Ridge regression, neural networks,
    and random forests.
  \item Bryan Keller (Columbia): Propensity score matching and CART summarization.
  \end{itemize}
\end{singlespacing}

All presenters, with the exception of Schuler, submitted a written
report (most with other coauthor(s)).  These
submissions are all included in this issue.


\newpage
\vskip 0.2in
\bibliography{main}

\end{document}